\documentclass[12pt]{article}
\begin{document}

\setlength{\textwidth}{15truecm}
\setlength{\textheight}{23cm}
\baselineskip=24pt

\bigskip

\centerline {\bf {Large-Scale Asymmetry of Rotation Curves in }}

\centerline {\bf {Lopsided Spiral Galaxies}}

\medskip

\medskip

\centerline {Chanda J. Jog}
      
\centerline  { Department of Physics, Indian Institute of Science}

\centerline { Bangalore 560012, India}

\centerline {email: cjjog@physics.iisc.ernet.in}

\newpage

\noindent{\bf Abstract}

Many spiral galaxies   show a large-scale  asymmetry with
a $cos \phi $ dependence in their 
rotation curves as well as in their morphology, such as M101 and NGC 628. 
We show that both these features
can be explained by the response of a galactic disk to an imposed lopsided 
halo potential.
A perturbation potential of $ 5\%$ is deduced for the
morphologically lopsided galaxies in the Rix \& Zaritsky (1995) sample. 
This is shown to result in a difference  of $10 \%$ or $\geq 20-30 km s^{-1}$  
in the rotation velocity on the two sides of the major axis.
Interestingly, the observed isophotal asymmetry in a typical
spiral galaxy is not much smaller and it results in a velocity
asymmetry of $7 \%$ or $\sim$ 14-21 $km s^{-1}$. Hence, we predict that most 
spiral galaxies show a fairly significant rotational asymmetry. 
The rotation velocity is shown to be maximum along the elongated
isophote - 
in agreement with the observations along the SW in M101,
while it is minimum along the opposite direction. This result leads to the 
distinctive asymmetric shape of the 
rotation curve which rises more steeply in one half of the
galaxy than the other, as observed by Swaters et al. (1999). This shape is 
shown to be a robust feature and would result for any centrally concentrated 
disk. The net disk lopsidedness and hence the asymmetry in the rotation curve 
is predicted to increase with radius and hence can be best studied 
using HI gas as the tracer. 
 
Running Title : Asymmetric Rotation Curves in Lopsided Galaxies

\bigskip

\noindent {\bf Key Words:} Galaxies:kinematics and dynamics -Galaxies:ISM -
  Galaxies:spiral - Galaxies- structure - Galaxies: haloes - 
  Galaxies: rotation curves

\newpage

\noindent {\bf 1. Introduction } 

The rotation curve of a spiral galaxy measures the variation of
the rotation speed with radius in a galactic disk. This
kinematic information is of crucial importance in understanding
the structure and dynamics of spiral galaxies (e.g. Sofue \&
Rubin 2001). Typically, the rotation curve is assumed to be azimuthally 
symmetric for simplicity and this is used to obtain the corresponding 
axisymmetric mass distribution in a galaxy that is supported rotationally.
(e.g., Binney \& Tremaine 1987).

The observed rotation curves in spiral galaxies, however, often show a 
deviation from a smooth circular rotation w.r.t. the galactic centre. 
The local deviations of a few $km s^{-1}$ are 
well-known (Shane \& Bieger-Smith 1966) and are believed 
to be due to the streaming motions associated with spiral
features. What is not so well-appreciated is 
that the rotation curves in many spiral galaxies show large-scale or global 
asymmetry such that the shape and the maximum velocity are different 
in the two halves of a galaxy, as for example in M101 (Mihalas \& 
Binney 1980). This indicates an underlying {\it mass asymmetry} in the galaxy. 
The global asymmetry of rotation curves is not so widely recognized because 
the observational data are generally averaged out to present an artificially 
`axisymmetric' rotation curve so that the precious information on the weak 
azimuthal asymmetry in the observed rotation curve is lost. Hence only the 
highly asymmetric cases such as M101 are recognized as having a global 
asymmetry. Further, even when large rotational asymmetry is noted in 
other individual galaxies it has been attributed to arise not due to 
underlying mass asymmetry but to other processes such as,  due to absorption 
as in NGC 6181 (Burbidge, Burbidge, \& Prendergast 1965), or due to an 
ejection from the centre as in NGC 4088 (Carozzi-Meyssonnier 1978).

Interestingly, all the nearby galaxies where the kinematics can
be studied with good resolution, the rotation curve is observed
to be azimuthally  asymmetric, as for example in M31 (Simien et al. 1978),
M33 (Colin \& Athanassoula 1981), M81 (Rots 1975), and NGC 4321
(Knapen et al. 1993). Yet the fact that most galaxies show some asymmetry 
in their rotation curves is not generally well-known.
One of the aims of this paper is to highlight this point and to
stress that  observers should publish  the  rotation curve
along both the sides of the major axis, and if possible give the
full, azimuthal plot.

The observed large-scale asymmetry of rotation curves in the two halves of a 
galactic disk was first noted and underlined for a number of galaxies by 
Huchtmeier (1975), who showed that the difference in the rotational 
velocities could be  $\geq 20 km s^{-1}$. A study of
 CO rotation curves by Sofue (1996) also illustrates such asymmetry
in the inner part of the optical disk.
A recent study of a large sample confirms that the rotation curve 
asymmetry is the norm rather than the exception (Kannappan \& Fabricant 2001).
In addition, in many cases the shape of the rotation curve in the 
two halves of a galaxy is observed to be asymmetric, with the
curve rising faster in one half
of the galaxy than the other half (Sancisi 1996, Swaters et al. 1999).

Recent near-IR observations have also revealed  
morphological asymmetry in the underlying old stellar disks in a
large fraction of galaxies studied, and the 
magnitude of the various azimuthal Fourier components denoted by
$m$ has been measured  (Block et al. 1994, Rix \& Zaritsky 1995, 
Zaritsky \& Rix 1997, Rudnick \& Rix 1998), also see Conselice (1997), and  
Kornreich, Haynes \& Lovelace (1998) who measure the asymmetry
in an average way. The HI distribution also shows mass asymmetry
as seen in the global profiles (Richter \& Sancisi 1994, Haynes et al. 1998).
 The isophotal asymmetry indicates asymmetry in the underlying disk mass 
distribution. It has been proposed (e.g., Weinberg 1995, Jog 1997) that the 
mass asymmetry occurs as a disk repsonse to an imposed halo 
distortion, the latter arising say due to a galaxy interaction. 

The origin of the asymmetry in the rotation curves has not been addressed 
theoretically so far. In this paper, we show that the global asymmetry in the 
rotation curves as well as morphology can be  explained naturally as 
arising due to the response of stars and gas in a galactic disk to an imposed 
lopsided ($m=1$) halo potential.  We note that even a small lopsided 
perturbation potential results in highly disturbed kinematics 
(Weinberg 1995, Jog 1997), hence it is easy to detect kinematic signatures 
of lopsidedness, especially the asymmetry in the rotation curve. 
 From the calculation of perturbed orbits, we show that along the major axis 
of an orbit, the rotation velocity is a maximum along the maximum magnitude 
of the perturbation  potential and is a minimum along the 
opposite direction (Jog 1997).
Such an asymmetry of the rotation velocity along the major axis
of an orbit was also pointed out by Earn \& Lynden-Bell (1996), and Syer
\& Tremaine (1998), though these authors do not apply this result to the 
rotation curve in a galaxy. A similar asymmetry in the rotation curve  
resulting from an imposed $m=2$ potential was studied by 
Gerhard \& Vietri (1986).

In this paper, we obtain the typical velocity differences for galaxies for a 
lopsided potential the value of which is deduced from the observed 
morphological asymmetry, and compare these with observed
velocity asymmetry. We also obtain the results for 
the inverse case, namely from the observed velocity difference known for the 
few galaxies, we can deduce the lopsided potential giving rise to the 
rotational asymmetry.
Further, we  tie in the velocity asymmetry to the isophotal elongation. 
The maximum velocity is shown to be along the elongated isophotes, this 
result agrees with the observations of M101. This correlation also 
allows us to explain the asymmetric shapes of the rotation curves observed 
in lopsided galaxies.

Section 2 contains the derivation of the asymmetric rotation
curve for a flat rotation curve, and also for a general,
power-law rotation curve. Applications of results to galaxies are given in 
Section 3. A few general points are discussed in Section 4, and the results 
from this paper are summarised in Section 5.

\noindent{\bf 2. Asymmetry in Rotation Curve}

\noindent {\it 2.1 Orbits}

We use galactic cylindrical co-ordinates $R, \phi, $ and $z$.
We assume the galactic disk to be azimuthally symmetric 
on which a small lopsided halo potential is imposed. 
The unperturbed potential, $\psi_0 (R) $ is taken to have a
logarithmic form for simplicity, and because it describes
the region of flat rotation curve:

$$ \psi_0 (R) = \: {V_c}^2 \: \: ln \: R  \eqno (1) $$

\noindent where $V_c$ is the constant rotational velocity. The
perturbation potential $\psi_{lop} (R)$ is taken to be
non-rotating and of the following form:

$$ \psi_{lop} (R, \phi) = \: {V_c}^2 \: \epsilon_{lop} \: cos \: \phi 
                     \eqno (2) $$

\noindent where $\epsilon_{lop}$ is a small, constant dimensionless parameter,
denoting the perturbation in the potential. The case of a
rotating perturbation potential is treated in the Appendix A.

Using the first-order epicyclic theory, the equations of motion
for $R$, the orbital radius, and $V_{\phi}$,  the azimuthal
component of velocity for the perturbed closed
orbits around $R_0, \phi_0$ are obtained to be (Jog 2000, Appendix A)
respectively:

$$ R \: = \: R_0 \: ( 1 - 2 \: \epsilon_{lop} \: cos \phi_0)  \eqno (3) $$
 
$$ V_{\phi} \:  = \: V_c \: ( 1 \: +  \: \epsilon_{lop} \: cos \phi_0) 
                        \eqno (4) $$

Thus, the orbit is shortened along $\phi = 0^o$ where the
perturbation potential is a maximum (see eq.[2]), and it is
elongated along the opposite direction along the major axis,
along $\phi = 180^o $. Note that
 $\Delta V_{\phi}$, the velocity difference at the two ends of
the major axis ($\phi = 0^o$ and $\phi = 180^o$, respectively) is
equal to:

$$ \Delta V_{\phi} = 2 V_c \epsilon_{lop} \eqno (5) $$

{\bf Thus the $\%$ velocity difference is exactly twice that of the 
perturbation in the potential.}

\noindent {\it 2.2 Isophotes}

The mass distribution in a spiral galactic disk is observed to
fall exponentially with radius (Freeman 1970):

$$ \mu_{un} (R) = \mu_0 \: exp (- \frac {R}{R_{exp}}) \eqno (6) $$

\noindent where $\mu_0$ is the central extrapolated surface density and
$R_{exp}$   is the scale length of the exponential disk. For the
perturbed case, the resulting effective surface density of the
perturbed orbits in an exponential disk is defined by (see Rix
\& Zaritsky 1995, Jog 1997) to be:

$$ \mu (R, \phi) = \mu_0 \: exp \left [ - \frac {R}{R_{exp}} \left(
        1 \: - \: \frac {\epsilon_{iso}}{2} \: cos \phi \right) \right]
	  \eqno (7) $$

An isophote is defined to be a curve of constant intensity or surface density
for a constant mass-to-light ratio. For a particular isophote,
the term in brackets in equation (7) is a constant, and defines
the parametric form of the isophote. Thus, the minimum radius of an
isophote occurs along $\phi = 180^o$, while the maximum extent
occurs along the $\phi = 0^o$. Physically this is because the 
surface density falls off exponentially with radius and hence an isophote
 must be elongated in regions which show local density enhancement 
  (that is, along $\phi = 0^o $, see eq.[7]). This correlation is true
 for any centrally concentrated disk.

Note that the elongation in an isophote is opposite to the behaviour 
of an individual orbit (Section 2.1). {\bf The rotation velocity is a maximum 
along the short side  of an orbit}, that is, along $\phi = 0^0$ 
(see eqns.[3]-[4]),
{\bf that is along the elongated side of the isophote}. The last point is
somewhat counter-intuitive and arises due to the self-gravity in
the disk, and we show that it agrees with
observations (Section 3.4).

The isophotal shapes for an exponential disk are given following
the procedure as in Jog (1997), and Jog (2000). For such a disk,
 $A_1 / A_0$, the fractional amplitude
of the $m=1$ azimuthal Fourier component of the surface
brightness is obtained to be:

$$ A_1 / A_0 \: = \: \vert - \frac {\Delta R}{R} \: \frac
{R}{R_{exp}} \vert                      \eqno (8) $$

\noindent where  $\Delta R / R$ is 
the distortion in the isophote. The amplitude $A_1/ A_0$ 
 is related to $\epsilon_{iso}$, the ellipticity of the
isophote at $R$, as follows: 

$$ A_1 / A_0 \: = \: \frac {\epsilon_{iso}}{2} \: \frac {R}{R_{exp}}
                         \eqno (9) $$
  
Because of the orbital velocity changes along an orbit, the
associated surface density also changes as a function of the
angle $\phi$. By solving together the equations of perturbed
motion, the continuity equation, and the effective surface
density (eq. [7]),
we obtain the following relation between the perturbation parameter
$\epsilon_{lop}$ in the potential, and the resulting asymmetry in the 
isophotes as denoted by $\epsilon_{iso}$ or by $A_1 /A_0$ (see
Jog 2000, Appendix A) to be:

$$ \epsilon_{iso} / \epsilon_{lop} \: = \: 4 ( 1 - 
         \frac {1}{2} \frac {R_{exp}}{R})           \eqno (10) $$

\noindent and,

$$ \epsilon_{lop}  =  \frac {A_1 /A_0}  {(\frac{2R}{R_{exp}}) - 1}
                         \eqno (11) $$

\noindent where the $R \geq R_{exp}$ since the calculation of
loop orbits is valid for this range only (Rix \& Zaritsky 1995).
The above equations are valid for a self-consistent disk
response for the lopsided ($m =1$) perturbation halo component,
which takes account of the negative disk response due to the
disk self-gravity  to the imposed potential.
Thus the $\epsilon_{lop}$ obtained in equation (11) is the net
lopsided perturbation parameter that affects the disk, which is
smaller by a reduction factor ($< 1$) compared to the 
`original' value for the halo (Jog 1999, Jog 2000).

\noindent {\it 2.3 Orbits and Isophotes in a Power-law Rotation Curve}

We obtain the results for closed orbits and isophotes in a
lopsided perturbation potential for an exponential disk which
obeys a general, power-law rotation curve given by

$$ V = V_c \: (R/ R_0)^{\alpha}   \eqno (12) $$

\noindent where $V_c$ is the azimuthal velocity at $R_0$, and
$\alpha$ is a non-zero small number ($ < 1 $) and is the
logarithmic slope of the rotation curve. The corresponding 
unperturbed potential can be obtained to be (e.g., Kuijken \&
Tremaine 1994):

$$ \psi_0 (R) = (V_c ^2 / 2 \alpha) \: (R/ R_0)^{2 \alpha} 
           \eqno (13) $$

\noindent In this case, the epicyclic frequency, $\kappa$ is
given by ${\kappa}^2 / {\Omega}^2 = 2 ( 1 + \alpha)$. In analogy
with the case of the flat rotation curve (eq.[2]), we assume the
perturbation halo potential in this case to be

$$ \psi_{lop} (R, \phi) = [ V_c (R/ R_0)^{\alpha}]^2 \: \epsilon_{lop} \:
       cos \phi   \eqno (14) $$

Using the first-order epicyclic theory as in Jog (2000), we
obtain the equations of motion for closed perturbed orbits
around $R_0, \phi_0$ in this case to be:

$$ R = R_0 \: \left [ 1 - 2 \epsilon_{lop} \left ( \frac {1 + \alpha}
      {1 + 2 \alpha} \right ) cos \phi_0  \right ]    \eqno (15) $$

$$ V_{\phi}  = V_c  \: \left [ 1 + \epsilon_{lop} \left (\frac {1}
      {1 + 2 \alpha} \right) cos \phi_0  \right ]    \eqno (16) $$

Hence, $\Delta V_{\phi}$ , the difference on the two sides of
the major axis in this case is:

$$ \Delta V_{\phi}  = 2 V_c \epsilon_{lop} \left (\frac {1}
      {1 + 2 \alpha} \right ) \eqno (17) $$
      
On comparing this with equation (5), it can be seen that the \%
velocity difference on the two ends of the major axis is smaller
in this case by a factor of $ 1 / (1 + 2 \alpha) $ than the
value for the flat rotation curve.

The rotational velocity is a maximum at $\phi = 0^o$ (eq.[16])
which is along the minimum of the orbital radius (eq.[15)). Hence
following the discussion as in Section 2.2 it can be seen that
in this case also the maximum rotational velocity will be along
the elongated part of an isophote.

Following an analysis similar to that is Section 2.2, it can be
shown that the ratio of the ellipticity of an isophote,
$\epsilon_{iso}$, to $\epsilon_{lop}$ in this case is given by:

$$ \frac {\epsilon_{iso}}{\epsilon_{lop}} \: = \: 4 \: \left [ \left (\frac
{1 + \alpha}{1 + 2 \alpha} \right) \: - \: \frac {R_{exp}}{2 R}
\right] \eqno (18) $$

Using the relation between $\epsilon_{iso}$ and the lopsided amplitude
$A_1 / A_0$ (eq.[9]), we get in this case:       

$$ \epsilon_{lop} \: = \: \frac {A_1 /A_0}{\left[ (2R / R_{exp})
\left (\frac  {1 + \alpha}{1 + 2 \alpha} \right) - 1 \right] }
     \eqno (19)  $$

Check that in the limit of $\alpha = 0 $, the last two equations
reduce to the equations (10) and (11) respectively which are valid for
a flat rotation curve, as expected.

\noindent {\bf 3. Results}

\noindent {\it 3.1 Resulting Asymmetry in Rotation Velocity, $\Delta V_{\phi}$ }

From the observed isophotal disk asymmetry  values, $A_1 /A_0$,  at a
radius 2.5 $R_{exp}$ (Rix \& Zaritsky 1995) , we
deduce the perturbation parameter $\epsilon_{lop}$ (eq.[11]) for
the lopsided potential, and thus obtain the resulting asymmetry
in the rotation velocity (eq.[5]).
{\it We show below that even a small perturbation potential gives rise to a
large kinematic asymmetry, hence it is easy to detect.}
For the typical morphologically lopsided galaxies in the Rix \& Zaritsky 
(1995) sample,
with $A_1 / A_0 \geq 0.2$, the perturbation parameter
$\epsilon_{lop}$ is obtained to be $\geq 0.05$ and the net velocity 
asymmetry $\Delta V_{\phi} \geq 0.10 V_c$, or $\geq 10 \%$ of the maximum 
rotation velocity. Thus for a typical range of maximum
rotational velocity of 200-300 $km s^{-1}$ for a giant spiral galaxy  
(Binney \& Tremaine 1987), the resulting magnitude of velocity
asymmetry lies in the range of $\geq$ 20-30 $km s^{-1}$. 
 Nearly a
quarter of the sample in Rix \& Zaritsky (1995) is strongly lopsided 
$A_1 / A_0 \geq 0.3$,
for which the velocity asymmetry is higher $\sim 30-45 km s^{-1}$.
This is closer to the value of the velocity asymmetry observed
in NGC 628 and in M101 (Kamphuis 1993).

The sample studied by Zaritsky \& Rix (1997) is larger, and shows
the same typical values for morphological lopsidedness. However, the value
of asymmetry given is an average over a radial range of 1.5-2.5 exponential
disk radii, and hence we cannot use equation (11) directly to obtain the
value of the lopsided potential for this sample.

An important point is that the average or typical value
of isophotal asymmetry for the Rix \& Zaritsky (1995) sample is 
 $A_1 / A_0 =  0.14$, which we note is not much smaller than the
value they use to define a lopsided
galaxy ($A_1 / A_0 \geq 0.2$). For this average asymmetry, the resulting
typical $\epsilon_{lop}$ is = 0.035 and the typical velocity difference 
is $ 7 \% $ of the maximum rotation velocity, or 
$14 - 21 km s^{-1}$. Thus we predict that most spiral galaxies
show a fairly significant rotational asymmetry, that can be
easily checked by future observations.

\noindent{\it 3.1.1 Comparison of $\Delta V_{\phi}$   with Observations:}

As a direct verification of our model, we consider the specific case of 
one galaxy, NGC 991, for which the observed values for both the asymmetry 
in morphology and that in kinematics are known in the same radial range.
We use the observed $A_1/ A_0$ amplitude 
values to obtain $\epsilon_{lop}$, and hence calculate the resulting 
rotation velocity asymmetry. We show that this agrees well with
the observed rotational asymmetry. This galaxy has a disk
scalelength of 19.5" (Rix \& Zaritsky 1995). The average
amplitude for lopsidedness between 1.5-2.5 disk scalelengths, or
at an average radius of 2 disk scalengths = 37" , is 0.224 
(Zaritsky \& Rix 1997).

The detailed kinematics for NGC 991 has been observed by
Kornreich et al. (2000), see their Figure 2. By fitting the observed 
rotation curve $V_c \: \: sin i $ between 10"-100" using equation (12) 
we get the slope $\alpha$ of the rising rotation curve to be = 0.36. 
Here $i$ is the inclination angle.
Using this, we apply equation (19) at a radius of 37", and thus
obtain $\epsilon_{lop}$, the lopsided perturbation parameter to be 
 $0.104$. Using equation (17), we predict the fractional difference or 
 asymmetry in the rotation curve on the two sides of the major axis to be
0.12. The observed rotational asymmetry at this average radius
is $\sim 5/35$ = 0.14. Note that this ratio is independent of the
inclination angle $i$. Thus the predicted value of rotational asymmetry
agrees to within $17 \% $ of the observed value, this supports 
our model.

     Unfortunately at this time such a detailed comparison is
not possible for more galaxies because of lack of near-IR
photometric data and detailed kinematic data for the same galaxies
(also see Section 3.2).

The results for the typical values of the rotation velocity asymmetry 
obtained (see Section 3.1) agree well with the average measurement of the 
rotation curve asymmetry of $4 - 10 \% $ for a sample of 9 Sa galaxies 
by Kornreich et al. (2001), which as per our model covers the range of weak to
typical lopsided galaxies. These galaxies show little
morphological asymmetry in the inner/ optical disk but show a
large kinematical asymmetry in HI in the outer region. In
contrast, the values for the rotation curve asymmetry are much higher
in the study of nine morphologically asymmetric galaxies by Kornreich
et al (2000), see their Table 4. In that work, 3 out of nine galaxies 
show strong asymmetry $(> 10 \% )$ in the rotation curve while the rest 
show weak asymmetry of $(< 10 \% )$. This confirms our picture
that the kinematic and morphological lopsidedness in a galactic disk
are causally related. 

Note that the above asymmetry is the maximum that can be observed when the 
viewing angle is the most favourable, namely along the minor axis of the 
galaxy $(90^o - 270^o)$, so that the rotation velocity at the ends of the 
major axis (see eq.[5]) is along the line of sight. The magnitude for any 
other orientation is 
smaller, and in the extreme opposite case when the line of sight is along 
the major axis, no asymmetry in rotation velocity is observed.

\noindent {\it 3.2 Lopsided Potential from observed {$\Delta V_{\phi}$} :
    inverse application}

The  observed velocity difference at the two ends of the major axis 
     yields a value for the perturbation parameter in
the halo potential $\epsilon_{lop}$ (see eq.[5]) withought the necessity of 
detailed mapping. This is a lower limit on the actual value of $\epsilon_{lop}$
because the observed $\Delta V_{\phi}$ would be smaller than
that given by equation (5) for any line of sight other than
along the minor axis, which as we argued above is the
most favourable for observing the velocity asymmetry.

This method is particularly important for the edge-on galaxies, like NGC 891, 
where the morphological  asymmetry cannot be measured due to 
severe dust
extinction, and the problem of unique de-convolution into non-axisymmetric 
features. This method is also applicable for inclined but large-angular size 
galaxies for which the measurement of optical asymmetry is time-consuming and 
has  not been attempted so far, such as M101 and NGC 628. This is despite the 
fact that such galaxies look highly disturbed in optical images. From the 
measured value of the velocity difference of $\sim 75 km s^{-1}$
beyond 10$'$ or 3$R_{exp}$ for M101 (Kamphuis 1993), we obtain
(using eq.[5]) the 
 lopsidedness in the halo potential, $\epsilon_{lop}$ to
be = 0.2. This is about 4 times larger than the value  of  0.05 we derived
 (Section 3.1) for a typical morphologically lopsided galaxy 
 in the Rix \& Zaritsky (1995) sample. Similarly, for the observed 
 velocity difference across the major axis of $\sim 20-30 km s^{-1}$ for 
 NGC 628 (Kamphuis 1993), at a constant velocity of $200$ km s$^{-1}$, we 
 obtain $\epsilon_{lop} \sim 0.05 - 0.075$.
These are upper limits on the actual asymmetry in the potential
of both these galaxies
since both these show strong evidence for gas infall at large radii
which could cause kinematic asymmetry due to reasons other than
the disk response to a lopsided halo potential.

An interesting application of this idea of obtaining the
lopsidedness in the halo
potential from the observed velocity asymmetry is for the case of
the Milky Way.
 It is well-known that the rotation curves in the northern and
southern hemispheres in the Galaxy show a global asymmetry and the diference
in the two values is $\sim 8 km s^{-1}$ (Kerr 1964).
Following the discussion in Section 3.1, it is shown below that
 if the observed global
asymmetry in the rotation curve of $8 km s^{-1}$ is to be
attributed to the disk response to a lopsided halo potential,
then our Galaxy is weakly lopsided. For the observed range of velocities
of 230 to 260 km s$^{-1}$ between a radius of 4 to 8 kpc (Kerr 1964),
and applying equation (12), we get the power-law index for the rotation curve 
to be 0.32 . Hence for the observed velocity difference of 8 km s$^{-1}$
and $V_c$ = 260   km s$^{-1}$, equation (17) gives $\epsilon_{lop}$
= 0.02. This is less than
half the value of $\sim 5 \% $ derived for the typical lopsided galaxy in 
the Rix \& Zaritsky (1995) sample (see Section 3.1).
The origin of the north/south rotation curve asymmetry in the Galaxy
was  first conjectured by Sancisi (1981) to be due to a lopsided
distribution, though in a qualitative way.

The above inverse problem was done more rigorously from the full
two-dimensional kinematic information for two galaxies by Schoenmakers, Franx,
\& de Zeeuw (1997).
However the various Fourier ($m$) components are coupled and it is
not possible to get a unique value of lopsided parameter for the
potential from such an analysis.
Our method of measuring asymmetry in the rotation curve has the
benefit of being a very simple measurement, and is especially
applicable for edge-on galaxies where the decomposition of
kinematic data would be hard.

It would be worth comparing the value of the $\epsilon_{lop}$
as obtained from the rotation curve  and that from the
near-IR isophotal analysis. If these agree, then that would
support our model. Unfortunately, the galaxies in the Rix \& Zaritsky
(1995) sample are chosen to be of small-angular size to allow a
single CCD frame measurement, and only a few of these have
kinematic data in the literature. Conversely, the nearby, large-angular size 
galaxies for which rotational asymmetry is studied such as M101 do 
not have measured isophotal asymmetry values in the near-IR.

\noindent {\it 3.3 Dependence of Asymmetry on Radius and Tracer}

The magnitude of the lopsided distribution in old stars, $A_1/A_0$, is
observed to be important beyond 1.5 exponential disk radii and
it increases with radius (Rix \& Zaritsky 1995). 
The signal-to-noise ratio limited their measuements to only 2.5 exponential 
disk radii. The lopsided disk distribution was first observed in HI, at radii
several times farther out than the optical disk as in M101, and
IC 342 (Baldwin, Lynden-Bell, \& Sancisi 1980), although the amplitudes 
of lopsidedness in HI distribution were not measured.
Theoretical work has shown (Jog 1999) that the disk self-gravity resists the 
imposed lopsided gravitational field, and hence the net, self-consistent 
lopsided distribution is only important in the outer region of a galactic 
disk, beyond two exponential disk radii (Jog 2000) indicating
the increasing dynamical importance of the halo compared to the disk
at large radii.  This agrees reasonably well with the near-IR 
observations of Rix \& Zaritsky (1995), and 
 Zaritsky \& Rix (1997) - the latter give an average value of the
lopsided amplitudes between 1.5-2.5 exponential disk radii. 
The large kinematical asymmetry in HI in the outer region of
galaxies that may show little morphological asymmetry in the inner or optical 
disk as observed by Kornreich et al (2000) (see Section 3.1.1) also 
agrees with this prediction.

Thus, we note that the HI component can act as an excellent
 diagnostic of the halo lopsidedness, since it typically
extends 2-3 times beyond the optical disk (e.g. Giovanelli \& Haynes 1988).
Hence we predict that the asymmetric rotation curve in a typical spiral galaxy
will be most easily detected in the outer galactic disk, studied using HI as 
a tracer.

\noindent {\it 3.4 Correlation of Velocity with Isophotal orientation}

In our model, the disk responds to the halo distortion, and hence
the asymmetry in the rotation velocity and the isophotal elongation have 
a unique correlation such that  the maximum rotation velocity is
along the elongated side of the isophote as shown in Section 2.2.
 This prediction agrees with the
rotation curve in M101 which shows a maximum velocity along the
southwest (Kamphuis 1993) along which the surface density is
also higher, and along which the isophotes in the inner galaxy
are elongated. This is also true for the case of IC342 which has a
high surface density along the NW (Baldwin et al. 1980), and the
rotation curve along the W is higher by $\sim 20 km s^{-1}$
than along the E (Sofue 1996). In order for the above
correlation to be detected, it is implicitly assumed that
the tracer exists upto the larger extent on the elongated side
of the isophote so that the above asymmetry is manifested.
The above correlation between the maximum of the rotation velocity 
and the isophotal elongation is also valid for a general, power-law
rotation curve (Section 2.3), and for a perturbation potential
with a non-zero pattern speed (Appendix A).

The orbital and isophotal elongation
occur along the opposite directions of the major axis, as shown
in Section 2.2. Since the observed quantity is the isophote rather than an
orbit, we have given the above correlation between the maximum
velocity and the elongation in the isophote.
In contrast, the models for disk lopsidedness by Earn \&
Lynden-Bell (1996) and by Noordermeer, Sparke, \& Levine (2001), do not make 
the distinction between the elongation in the 
orbit and isophotes. They also show that the the maximum velocity 
is along the "small side" of the orbit but they do not make any 
predictions about the maximum velocity and the isophotal orientation, 
which is the only quantity that may be directly verified from obsevations.

We note that some galaxies show the reverse result where the
maximum velocity is seen along the less elongated side of the
isophote such as along the NW in NGC 4565, or along the side with lower 
surface density such as along NW in IC 342 (Sofue 1996). Our picture cannot 
explain these 
anomalous cases. Perhaps gas infall or a central bar could be responsible in 
severely disturbing the kinematics in these cases.

In some galaxies, there is an additional complication namely the
sense of elongation is in opposite directions in the inner
and the outer
regions of the galaxy. This is true for example in M101
where the isophotes in optical or
isodensity contours for HI are elongated along
the SW in the inner region, while in the outer region the HI
contours are elongated in the opposite direction, namely the
NE. This can be explained naturally in our model of a galactic disk
respoding to an imposed halo potential, which is probably
generated via galaxy encounters. This is because the subsequent galaxy
 encounters are un-correlated, therefore the lopsidedness  in different 
the radial regions of a galaxy halo and hence the net disk response may 
show different orientation as argued by Jog (1997). In this
case, the rotation curve on one side of a major axis would change from
being a maximum at a lower radius to a minimum at a higher radius, and 
hence we argue that the resulting
rotation curve in this case would have a `braided' appearance.
It would be interesting to check this with the full data 
covering all azimuthal points are measured in future, say for M101.

The above discussion on the correlation between isophotes and
rotation velocity
 is  applicable to galaxies with nested, oval
or egg-shaped isophotes such as M101 which arise due to a constant phase 
with radius of the halo potential as shown by Jog (1997). In contrast, the 
galaxies with a strong radial dependence of phase give one-armed spirals
such as M51 or NGC 4254. In the latter case the velocity will be
maximum along different azimuthal angles at different radii
since the major axis orientation changes with radius. Here too
the maximum of the surface density would coincide with the
elongation in the arm and we expect the velocity to be a maximum
along the longer and more prominent arm, as along the NE in M51. Unfortunately,
only the average rotation curves for galaxies are given in the literature.
We urge the observers to publish full, 2-D rotation curves which
will allow a comparison of the above prediction with observations. 

\noindent {\it 3.5 Asymmetric Shapes of Rotation Curves}

Recent detailed kinematical study of HI distribution in galaxies
shows distinctive shapes of rotation curves in the two halves, such that 
in one half of the galaxy the rotation curve rises slowly than in the other 
half, and reaches the flat part at a larger radius (Sancisi 1996, Swaters 
et al. 1999). In retrospect, this asymmetry in shape is also seen
in the earlier work where the rotation curve on the two sides of a
major axis was measured (e.g. Huchtmeier 1975, Sofue 1996).
We also note that the slow rising part is observed to end in a higher value 
of the flat rotation curve, and we argue below that this can be explained 
physically in terms of the response of a centrally concentrated galactic disk.

 As shown in Section 2.3, for an exponential disk with a rising
rotation curve, the maximum rotation 
 velocity occurs along the direction of elongataion of an isophote. 
 Thus, the maximum velocity will be reached at a larger radial distance 
 from the galactic centre. The rotation curve would therefore 
 increase gradually and have  a smaller slope on this side of the
major axis, where the rotational velocity is higher. Thus we can give 
a physical explanation as to why the slow
rising part of a rotation curve ends in a higher value of the flat rotation 
velocity as observed by Swaters et al. (1999), and it occurs at
a larger radius in the disk.
Since the correlation between the maximum velocity and the sense
of elongation of an isophote is valid for any centrally concentrated
disk (see Section 2.2), hence the above argument about the shape
of the rotation curve is valid for any realistic, centrally
concentrated disk.

\noindent {\bf 4. Discussion}

\noindent {\it 4.1 Asymmetry in galaxies in groups}

In our model, the  asymmetry in a galactic disk results from the
self-consistent disk response to a halo distortion. The latter is
most likely to arise in a tidal enclounter between galaxies as
shown by the work of Weinberg (1995). The galaxies in a group
undergo more frequent encounters than the field case and with a
similar relative velocity, hence we expect a large fraction of
galaxies in groups to exhibit lopsided distribution. 
This prediction is confirmed from the  observations on
Hickson group galaxies (Rubin, Hunter, \& Ford 1991)
where  $\geq 50 \%$ of galaxies show lopsided rotation curves. This
is much larger than the case of field spiral galaxies where only
$\sim 25 \%$ show asymmetric rotation curves (Rubin, Waterman, \&
Kenney 1999, Sofue \& Rubin 2001). The observations of rotation curves of 
30 galaxies in 20 Hickson compact groups by Nishiura et al. (2001) confirms
the above result that asymmetric rotation curves are more
frequently seen in these galaxies than in the field spiral galaxies.

A study of five major spirals in the nearby Sculptor group of galaxies
(Schoenmakers 2000) shows that all five show kinematic
lopsidedness while two are globally elongated or
are morphologically asymmetric. This is higher than the $\sim 30
\%$ of galaxies showing morphological lopsided distribution in the
field case (Rix \& Zaritsky 1995). This confirms our argument that the 
percentage of galaxies showing lopsided distribution is higher in groups.

In comparing the field versus the group cases, it should be
remembered that the adopted definition of what constitutes lopsidedness 
is somewhat arbitrary in each paper. However, since the
discrepancy between the field and the group cases is large, it
probably points to a genuine difference between the two cases.

\noindent {\it 4.2 Morphological vs. Kinematical Lopsidedness}

There has been some discussion in the literature as to whether 
the observed morphological and kinematical lopsidedness in a disk are
correlated or not. Richter \& Sancisi (1994) argue that the two
are correlated except for exceptional cases like NGC 4395.
This dwarf galaxy shows no asymmetry in morphology in
stars and HI but does show a kinematical lopsidedness 
(Swaters et al. 1999). The reverse is
also true for a number of dwarf galaxies which may show no
kinematic asymmetry but do show morphological asymmetry (Swaters
 1999). We note that these are dwarf galaxies, and it is
possible that for these galaxies the model of an off-centre disk in
a halo with retrograde orbits (Noordermeer et al. 2001)
may apply, where the two forms of lopsidedness may not be correlated. 
This model requires the disk to be within the flat
density, central core of the halo and hence is valid for late-type dwarf 
galaxies only. We note that in their model, the resulting disk lopsided 
distribution is only seen in the inner two disk-scale lengths, hence 
their model is not valid for the vast majority of large spiral
galaxies which show lopsided distribution in HI at radii far
outside the optical region (Baldwin et al. 1980) - that is, at several times 
the disk scale-length. Their model also cannot explain the increasing disk 
lopsidedness in stars beyond 2 disk scale-lengths as observed for giant 
spirals by Rix \& Zaritsky (1995).

Kornreich et al (2000)  have observed nine face-on, giant spiral galaxies and 
have  argued that the different indicators of morphological and
kinematical asymmetry are not well-correlated, except
that the morphological asymmetry is correlated with the asymmetry
in the position angle differences. The lack of correlation could
be due to the fact that their indicators of lopsidedness denote
average quantities. Further, their definition of morphological
asymmetry (as developed by Kornreich et al 1998) is not correct
since it would consider
a one-armed spiral such as NGC 2326 to be symmetric, whereas Jog
(1997) has shown these to be asymmetric with a phase that varies
with the radius. Thus some of the galaxies they classify as
having morphological symmetry are actually asymmetric. This
could partly explain the discrepancy they observe.

The origin of lopsidedness in the few giant giant 
spirals which show little optical asymmetry but show kinematical asymmetry in
HI, and have counter-rotating cores such as NGC 4138 (Kornreich
et al. 2001), could be attributed to the counter-rotation as in
the model for lopsidedness developed by Sellwood \& Merritt (1994).
 Or this behaviour could be explained  by the fact 
that HI being at a larger radius is a better
tracer of lopsided distribution than the stars in the inner 
disk since the net disk lopsidedness increases with radius in
our model (see Section 3.3).

Physically it is hard to see how in general the two measures of 
lopsidedness could be 
un-correlated in a giant spiral galaxy since the two would be
coupled via the continuity eqaution (see Jog 1997). However, in some
cases, where there is another source of kinematic disturbance as
say due to gas infall at large radii (van der Hulst \& Sancisi 1988),
the gas kinematics may be disturbed without affecting the mass distribution
significantly. This is why the lopsided potential derived from
the observed velocity difference in M101 (Section 3.2) gives an
upper bound to the actual value.
 
Finally, we point out a couple of caveats that need to be kept in mind when 
measuring the asymmetry. First, in
 order to see if the rotation curve is symmetric or not, it
must be compared upto the same radius and in the same tracer on
both the sides of a major axis. When this is not done, it could lead to a 
spurious claim that the rotation is symmetric, as Noordermeer et al. (2001) 
claim to be true for NGC 891. Actually the rotation curve is measured upto 
different radii on the two sides of the major axis (Swaters, Sancisi \& van
der Hulst 1997), and hence is definitely asymmetric on a global scale. 
Upto the last common radius on the two sides, the rotation curve however is 
symmetric. Similarly, in NGC 6946, the HI extends much farther out (to 25
\% larger radius) along the NE along the major axis than the SW,
hence the rotation curve can only be defined upto a smaller
radius along the SW (Carignan et al. 1990). Hence, globally the
rotation curve is asymmetric in NGC 6946.

The other point is that, the net lopsidedness is predicted to be
higher at larger radii (Jog 1999), hence it is expected to be more easily
seen in HI since it extends out to a larger radius than do the
stars. This is in fact confirmed from the study of Kornreich et
al (2000), see Section 3.3.  Hence, when comparing the
lopsidedness of different galaxies, it is not correct to compare the 
asymmetry in stars measured upto a smaller radius with the asymmetry in HI 
measured upto a much larger radius.

\noindent {\bf 5. Conclusions}

 We obtain the self-consistent response of an exponential galactic disk to 
 an imposed lopsided halo potential, and show that the disk exhibits both 
morphological lopsidedness and an asymmetric rotation curve. We
study the case of a disk with a flat rotation curve, as well as a
disk with a general, power-law rotation curve. The
main results obtained are:

1. The \% velocity asymmetry in the rotation curve is twice that
of the perturbation in the potential. Hence, even a small perturbation 
potential of a few \% results in  a significant
asymmetry in the rotation curve. 

2.  From the observed isophotal asymmetry, it is predicted that the 
{\it typical} spiral galaxies will show an asymmetry in the rotation velocity 
of $ 14 - 21 km s^{-1}$ which should be easily detectable. 

For NGC 991, the calculated velocity asymmetry from the observed
isophotal lopsidedness is shown to agree fairly well with the observed 
asymmetry in the rotation curve, this confirms our model.

We show that if the well-known North/South asymmetry of the rotation 
curve of the Galaxy is attributed to the disk response to a halo perturbation, 
then the Galaxy is a weakly lopsided galaxy with a
perturbation potential of $2 \%$. 

3. The rotation velocity is shown to be a maximum along the elongated
isophote, in agreement with the observation for example 
along the SW in M101. Using this correlation, we can explain the
observed, asymmetric shape of the rotation curves is galaxies in
a natural way.

4. A tracer at larger radii, such as HI gas, will show a larger
rotational asymmetry, and hence will be a good diagnostic of the
lopsided halo potential.

\medskip

   We hope the present paper will motivate future observational
papers to give full azimuthal rotation data in galaxies, which will 
lead to a better understanding of galaxy asymmetry.

\medskip

\noindent {\bf Acknowledgements}

I would like to thank the referee for many constructive comments
which have greatly improved the paper, and also led to the addition of
the case of a general power-law rotation curve, and also the 
case of a rotating potential.

\newpage

\noindent {\bf Appendix A: Perturbation Halo Potential with a Non-zero
  Pattern Speed}

\medskip

Here we derive the equations for the closed orbits and the
isophotes in an exponential disk perturbed by a small lopsided potential
$\psi_{lop}$ which has a small, non-zero, constant pattern 
speed of rotation $\Omega_p$.  The unperturbed disk potential
is denoted by $\psi_0 (R) $.
The perturbation potential can be written as:

$$ \psi_{lop} (R, \phi) = \psi_{pert} (R) \: cos \phi  \eqno (A1) $$

\noindent where $\psi_{pert} (R)$ is the magnitude of the
lopsided perturbation potential. 

A perturbed orbit around the initial circular orbit at $R_0$ may
be written as: $R = R_0 + \delta R$ and $\phi = \phi_0 + \delta \phi$.
From the equation of motion in the
rotating frame for an unperturbed disk (Binney \& Tremaine 1987,
Chapter 3.3), we get

$$ \left (\frac {d \psi_0}{d R} \right )_{R_0} \: = \: R_0 \:
(d\phi_0 /dt + \Omega_p)^2     \eqno (A2) $$

For a disk in a rotational equilibrium, the angular velocity
$\Omega_0$ at $R_0$ is given by:

$$ \Omega_0 \: = \: \left [ \frac {1}{R_0} \left ( \frac {d \psi_0}
  {d R} \right )_{R_0} \: \right]^{1/2}    \eqno (A3) $$
  
Hence the angular velocity of the guiding centre at $R_0$ in the
rotating frame is given by  $ d\phi_0 /dt = \Omega_0 - \Omega_p$ .
On integrating with time, and appropriately choosing $t=0$, we get
$ \phi_0 = (\Omega_0 - \Omega_p) t $.    

Following the treatment for the first-order epicyclic theory for
a rotating frame with a small, non-zero pattern speed from
Binney \& Tremaine (1987), we get the following coupled equations
of motion for $\delta R$ and $\delta \phi$  :

$$ \frac {d^2 \delta R} {d t^2} = \: - \: \delta R \: \: \left(
3 {\Omega_0}^2 \: + \: \frac {{d}^2 \psi_0} { d
R^2}   \: \: _{(at R_0)}  \: \right ) \: \: $$

$$\: \: \: \: \: \: \:  - \: \: 
\left ( \frac { \Omega_0}{\Omega_0 - \Omega_p  } \frac {2 \psi_{pert} (R_0)}
{R_0} \: + \: 
 \frac {d  \psi_{pert}} {d R} \: \: _{(at R_0)}  \: \right)
 \: cos (\Omega_0 - \Omega_p ) t            \eqno (A4) $$ 

and,

$$ R_0 \frac {d^2  \delta \phi} {d t^2} \: + \: 2 \Omega_0  \: 
   \frac { d  \delta R} {d t} \:
   = \:  \frac {\psi_{pert} (R_0)} {R_0} \: sin (\Omega_0 - \Omega_p) t   
   \eqno (A5) $$
   
The first term in the parantheses on the right hand side of equation (A4) is 
the square of the standard first order epicyclic frequency, $\kappa$,  at 
 $R_0$. From the theory of a driven oscillator (e.g., Symon
1960), these may be solved to give the following solutions for the closed 
orbits in the rest frame of the rotating perturbation potential:

$$ R \: = \: R_0 \: - \: \frac { \left (  \frac 
 { \Omega_0}{\Omega_0 - \Omega_p  } \frac {2 \psi_{pert} (R_0)} {R_0}
  \: + \: (\frac {d  \psi_{pert} } { d R})_{(at R_0)}  \: 
  \right) \:  cos (\Omega_0 - \Omega_p) t}
  { {\kappa}^2 - (\Omega_0 - {\Omega_p})^2 }         \eqno (A6) $$

Hence, $V_R$, the perturbed radial velocity along this orbit is
given by

$$ V_R \: = \: \frac { \left (  \frac 
 { 2 \Omega_0 \psi_{pert} (R_0)} {R_0}
  \: + \: (\frac {d  \psi_{pert} } { d R})_{(at R_0)}  \: 
  \right) \:  sin (\Omega_0 - \Omega_p) t}
  { {\kappa}^2 - (\Omega_0 - {\Omega_p})^2 }         \eqno (A7) $$

The net azimuthal velocity, $V_{\phi}$, is given by:

$$ V_{\phi} \: = V_c \: + \: \frac {\Omega_0  \: (\frac {d \psi_{pert}}
 { d R})_{(at R_0)}} {{\kappa}^2 - (\Omega_0 - {\Omega_p})^2 } \:
  cos (\Omega_0 - \Omega_p) t $$
  
$$\: \: \: \: \: \:  \: + \: \frac {\left ( 2 {\Omega_0}^2 - 
 [{\kappa}^2 - (\Omega_0 - {\Omega_p})^2] \right) \: \psi_{pert} (R_0) }
 {R_0 (\Omega_0 - \Omega_p) [{\kappa}^2 - (\Omega_0 - {\Omega_p})^2]}
   \: cos (\Omega_0 - \Omega_p) t  \eqno (A8)  $$
  
Note that in the limit of $\Omega_p$ =0, and when $\psi_0 = V_c
^2 \: ln R$ and $\psi_{pert} =
V_c ^2 \: \epsilon_{lop}$, the above two equations reduce to the
equations for velocity for a non-rotating perturbation potential
as treated in Jog (2000), see their equation (A10).   
The maximum rotational velocity occurs along $(\Omega_0 -
\Omega_p) t = 0^o$, along which the perturbed
orbit has a minimum extent (see eq.[A6]). Hence, following the same
argument as in Section 2.2, it can be seen that in this case
also the rotation velocity is a maximum along the elongated side
of an isophote.

The above equation shows that a non-zero pattern speed
$\Omega_p$ does leave a signature in the resulting rotational
velocity that can be measured. In particular it can give rise
to resonances which are particularly important at higher radii where
$\Omega_0$ is low ($\sim \Omega_p$), and where lopsided
perturbation is more likely to occur (Jog 1999).
We note that the recent theoretical work on longevity of lopsided potential
argues that the pattern speed is likely to be low. In any case,
we have treated the case of a non-zero pattern speed here for the
sake of completeness.

We next obtain the isophotes for a specific case  for the sake of simplicity,
where the unpertubed and perturbed potential are taken to be respectively:

$$\psi_0 (R) \: = \:  V_c ^2 \: ln R   \eqno(A9)  $$

$$ \psi_{lop} (R, \phi) \: =  V_c ^2 \: \epsilon_{lop} \: 
   cos (\Omega_0 - \Omega_p) t    \eqno(A10) $$                      

These are written in analogy with the non-rotating case treated
in Section 2. We next apply the analysis as in Section 2 to the 
equations of motion obtained in this Appendix, and use the above
choice of potentials. This gives the following relation between
$\epsilon_{iso}$, the ellipticity of an isophote and
$\epsilon_{lop}$ :

$$ \frac {\epsilon_{iso}} {\epsilon_{lop}} \: = \: \frac {4 {{\Omega}_0}^2}
  {\kappa^2 - (\Omega_0 - \Omega_p)^2} \: \left [ 1 \: - \: \frac{R_{exp}}{R}
  \left( 1 - \frac {2 {\Omega_0}^2 - [{\kappa^2 - (\Omega_0 - \Omega_p)^2}]}
   {2 \: \Omega_0 \: (\Omega_0 - \Omega_p)} \right) \right]   \eqno(11) $$

Using the definition of $\epsilon_{iso}$ in terms
of the amplitude $A_1 / A_0$ (eq.[9]), this gives:

$$ \epsilon_{lop} \: = \: \frac {A_1 /A_0}{ \left [ \frac {2 {{\Omega}_0}^2}
  {\kappa^2 - (\Omega_0 - \Omega_p)^2} \right] \: 
  \left [ \frac{R}{R_{exp}} \: -
 \: \left( 1 - \frac {2 {\Omega_0}^2 - [{\kappa^2 - (\Omega_0 - \Omega_p)^2}]}
   {2 \: \Omega_0 \: (\Omega_0 - \Omega_p)}\right) \right]}   \eqno(12) $$

In the limit of $\Omega_p = 0$, the above two equations reduce
to equations (10) and (11) for the non-rotating potential, as expected.

\newpage

\noindent{\bf References}

\bigskip

\noindent Baldwin, J.E., Lynden-Bell, D., \& Sancisi, R. 1980, MNRAS,
           193, 313

\noindent Binney, J.,  \& Tremaine, S. 1987, Galactic Dynamics,
           (Princeton : Princeton University Press)

\noindent Block, D.L., Bertin, G., Stockton, A., Grosbol, P.,
  Moorwood, A.F.M.,  \& Peletier, R.F. 1994, A \& A, 288, 365

\noindent Burbidge, G., Bubidge, E., \& Prendergast, K.H. 1965,
   ApJ, 142, 641

\noindent Carignan, C., Charbonneau, P., Boulanger, F., \&
 Viallefond, F.  1990, A \& A, 234, 43

\noindent Carozzi-Meyssonnier, N. 1978, A \& A, 63, 415

\noindent Colin, J., \& Athanassoula, E. 1981, A \& A, 97, 63

\noindent Conselice, C.J. 1997, PASP, 109, 1251
 
\noindent Earn, D.J. D., \& Lynden-Bell, D. 1996, MNRAS, 278, 395

\noindent Freeman, K.C. 1970, ApJ, 160, 811
 
\noindent Gerhard, O.E.,  \& Vietri, M. 1986, MNRAS, 223, 377

\noindent Giovanelli, R. \& Haynes, M.P. 1988, in Galactic and
 Extragalactic radio Astronomy, eds. G.L. Verschuur and K.I. Kellermann 
 (2d ed.; New York: Springer), 522

\noindent Haynes, M.P., Hogg, D.E., Maddalena, R.J., Roberts,
 M.S., \& van Zee, L.  1998, AJ, 115, 62

\noindent Huchtmeier, W.K.  1975, A \& A, 45, 259 

\noindent Ideta, M. 2002, ApJ, 568, 190
 
\noindent Jog, C.J. 1997, ApJ, 488, 642

\noindent Jog, C.J. 1999, ApJ, 522, 661

\noindent Jog, C.J. 2000, ApJ, 542, 216

\noindent Kamphuis, J. 1993, University of Groningen, Ph.D. thesis.

\noindent Kannappan, S.J.  \& Fabricant, D. G. 2001, in Galaxy disks
 and Disk galaxies, ASP Conference series 230, eds. G. Jose, S.J. Furies, 
 M.C. Enrico (ASP: San Fransisco),  p. 449

\noindent Kerr, F.J. 1964, The Galaxy
and the Magellanic Clouds, IAU Symposium no. 20, eds. F.J. Kerr
\& Rodgers, A.W. (Canberra: Australian Acad. Sci.), 81

\noindent Knapen, J.K., Cepa, J., Beckman, J.E., del Rio, M.S.,
     \& Pedlar, A.    1993, ApJ, 416, 563
      
\noindent Kornreich, D.A., Haynes, M.P., \& Lovelace, R.V.E.
1998, AJ, 116, 2154

\noindent Kornreich, D.A., Haynes, M.P., Lovelace, R.V.E., \&
van Zee L. 2000, AJ, 120, 139

\noindent Kornreich, D.A., Haynes, M.P., Jore, K.P., \& Lovelace, R.V.E.
 2001, AJ, 121, 1358

\noindent Kuijken, K., \& Tremaine, S. 1994, ApJ, 421, 178 

\noindent Mihalas, D., \& Binney, J. 1980, Galactic Astronomy
  (San Francisco: Freeman)

\noindent Nishiura, S., Shimada, M., Ohyama, Y., Murayama, T.,
  \& Taniguchi, Y. 2000, AJ, 120, 1691

\noindent Noordermeer, E., Sparke, L.S., Levine, S.E. 2001,
MNRAS, 328. 1064

\noindent Richter O.-G., \& Sancisi, R. 1994, A \& A, 290, L9

\noindent Rix, H.-W., \& Zaritsky, D. 1995, ApJ, 447, 82 $\: \:$

\noindent Rots, A.H. 1975, A \& A, 45, 43

\noindent Rubin, V.C., Hunter, D.A., \& Ford, W.K., Jr.  1991, ApJS,
 76, 153

\noindent Rubin, V.C., Waterman, A.H., \& Kenney, J.D. 1999, AJ, 118, 236

\noindent Rudnick, G., \& Rix, H.-W. 1998, AJ, 116, 1163

\noindent Sancisi, R. 1981, in The Structure and Evolution of Normal
  Galaxies,  ed. S.M. Fall \& D. Lynden-Bell (Cambridge : Cambridge
  Univ. Press), 149

\noindent Sancisi, R. 1996, in New Light on Galaxy Evolution,
eds. R. Bender \& R.L. Davies (Dordrecht: Kluwer), p. 143.

\noindent Schoenmakers, R.H.M. 2000, in Small galaxy groups,ASP
conference series, Vol. 209 (ASP:San Fransisco), 54

\noindent Schoenmakers, R.H.M., Franx, M.  \& de Zeeuw, P.T. 1997, MNRAS, 
  292, 349

\noindent Sellwood, J.A., \& Merritt, D. 1994, ApJ, 425, 530

\noindent Shane, W.W., \& Bieger-Smith, G.P. 1966, BAN, 18, 263

\noindent Simien, F., Pellet, A., Monnet, G., Athanassoula, E.,
Maucherat, A., \& Courtes, G. 1978, A \& A, 67, 73

\noindent Sofue, Y. 1996, ApJ, 458, 120

\noindent Sofue, Y., \& Rubin, V.C. 2001, ARAA, 39, 137

\noindent Swaters, R.A. 1999, Ph.D. Thesis, University of Groningen.

\noindent Swaters, R.A., Sancisi, R., \& van der Hulst, J.M.
 1997, ApJ, 491, 140

\noindent Swaters, R.A., Schoenmakers, R.H.M., Sancisi, R., \&
Van Albada, T.S.  1999, MNRAS, 304, 330

\noindent Syer,D., \& Tremaine, S.D. 1996, MNRAS, 281, 925

\noindent Symon, K.R. 1960, Mechanics (Reading: Addison-Wesley)

\noindent Van der Hulst, J.M., \& Sancisi, R. 1988, AJ, 95, 1354

\noindent Weinberg, M.D. 1995, ApJ, 455, L31

\noindent   Zaritsky, D. \&  Rix, H.-W.  1997, ApJ, 477, 118

\end{document}